\documentclass[10pt]{article}

\usepackage{pslatex}
\usepackage{hyperref}
\usepackage{graphicx}

\makeatletter
\def\@sect#1#2#3#4#5#6[#7]#8{\ifnum #2>\c@secnumdepth
  \def\@svsec{}\else 
  \refstepcounter{#1}\edef\@svsec{\csname the#1\endcsname.\hskip0.5em}\fi
  \@tempskipa #5\relax
  \ifdim \@tempskipa>\z@
  \begingroup 
     #6\relax
     \@hangfrom{\hskip #3\relax\@svsec}{\interlinepenalty \@M #8\par}%
  \endgroup
  \csname #1mark\endcsname{#7}\addcontentsline
      {toc}{#1}{\ifnum #2>\c@secnumdepth \else
        \protect\numberline{\csname the#1\endcsname}\fi #7}%
  \else
    \def\@svsechd{#6\hskip #3\@svsec #8\csname #1mark\endcsname
      {#7}\addcontentsline{toc}{#1}{\ifnum #2>\c@secnumdepth \else
        \protect\numberline{\csname the#1\endcsname}\fi #7}}%
  \fi \@xsect{#5}}
\@addtoreset{equation}{section}
\@addtoreset{figure}{section}
\makeatother
\renewcommand\thesection{\Roman{section}}

\renewcommand\theequation{%
  \ifnum \value{section}>0
     \thesection.\arabic{equation}%
  \else
     \arabic{equation}%
  \fi}
\renewcommand\thefigure{%
  \ifnum \value{section}>0
     \thesection.\arabic{figure}%
  \else
     \arabic{figure}%
  \fi}

\def\EasyNData{\mbox{EasyNData}}
\begin{document}
\begin{center}
  {\Large\bf
   EasyNData\footnote{The program is available from
     http://puwer.web.cern.ch/puwer/EasyNData/}: 
   A simple tool to extract numerical values from published
   plots.
   }\\
  \vspace*{1cm}

   P. Uwer\footnote{Heisenberg Fellow}\\
  {\em Institut für Theoretische Teilchenphysik, 
        Universität Karlsruhe\\ 76128 Karlsruhe, Germany}\\
\end{center}
  \vspace*{0.5cm}

\centerline{\bf Abstract}
\begin{center}
  \parbox{0.8\textwidth}{
   The comparison of numerical data with published plots is a frequently 
   occurring task.
   In this article I present a short computer program written
   in Java\texttrademark\ helping in
   those cases where someone wants to get the numbers out of a plot but
   is not able to read the plot with a decent accuracy 
   and cannot contact the author of the
   plot directly for whatever reason. The accuracy reached by this
   method depends on many factors. For the examples illustrated in
   this paper a precision at the level of a few per mille 
   could be reached. 
   The tool might help in improving the quality of future publications.
}
\end{center}

\section{Introduction}
The comparison with published results which are only presented in form
of plots is a frequently occuring task. Usually one proceeds by first
looking at the plot and guessing whether the numbers might be the same as those
one has calculated. While this is sufficient for a rough estimate it is
clear that at a certain point one might want to make a more detailed
comparison. In that case the first thing to try is clearly to contact
the author of the plot and ask him politely for the raw data. Unfortunately in
many cases this does not work. It might be due to the fact that 
the person who did the plot left physics, or that simply the data files are 
somehow lost or that people just don't like to talk to each other. In these
cases one has to resort to other techniques. 
In the old days a good
technique was to print out the plot as large as possible, and then
equiped with a ruler to measure individual points which were afterwards
transformed using a calculator to get the value itself. 
While in principle nothing is
wrong with this technique it is tedious and error prone and somehow
does not qualify as a technique for the new millenium.
The uncertainty of points extracted in this way depends on the size of
the plot and on the ability to measure individual points. Using a
large format together with a ruler equiped with a magnifying glass a
decent accuracy can be reached. It should be clear that there is an
intrinsic limitation of the method due to the quality of the plot and
the line thickness. The obvious improvement of the method described 
above is to first
obtain a representation of the plot which can be displayed on the
screen and then using the mouse to read off individual data points.
After having calibrated once the mouse coordinates are translated 
automatically into the plot coordinates by a short program.  

\section{Program description}
The above mentioned method has been coded in a  
Java\texttrademark\ program called \EasyNData.
Most of it could be recycled from a much more complicated program
Map2GPS I wrote some time
ago in which points from scanned maps were translated to 
Universal Transverse Mercator (UTM) coordinates using the world
geodetic system of the year 1984 (WGS84). (The combination of the
ellipsoid WGS84 together with coordinate system UTM is used in most of the
modern GPS devices.) Having made this as a side remark I shall came back to
the description of the program \EasyNData\ presented in this
article.
The program is best described by a simple example.
Suppose 
we want to compare the data shown in Fig. 4 of
ref. \cite{Dittmaier:2007th} 
because we
did a second calculation and we do not want to contact the authors
directly\footnote{It is probably fair to say that a table with
  concrete
  values would have improved ref. \cite{Dittmaier:2007th}. Apparently
  this was somehow forgotten during the write-up.}. 
\begin{figure}[htbp]
  \begin{center}
    \leavevmode
    \includegraphics[width=10cm]{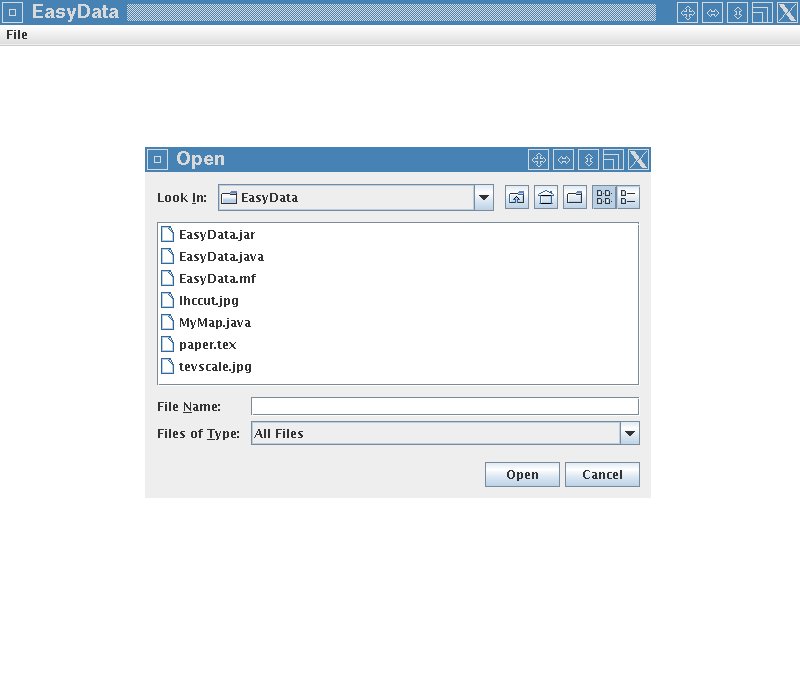}    
    \caption{File chooser to load the desired JPEG file.}
    \label{fig:0}
  \end{center}
\end{figure}
\begin{figure}[htbp]
  \begin{center}
    \leavevmode
    \includegraphics[width=10cm]{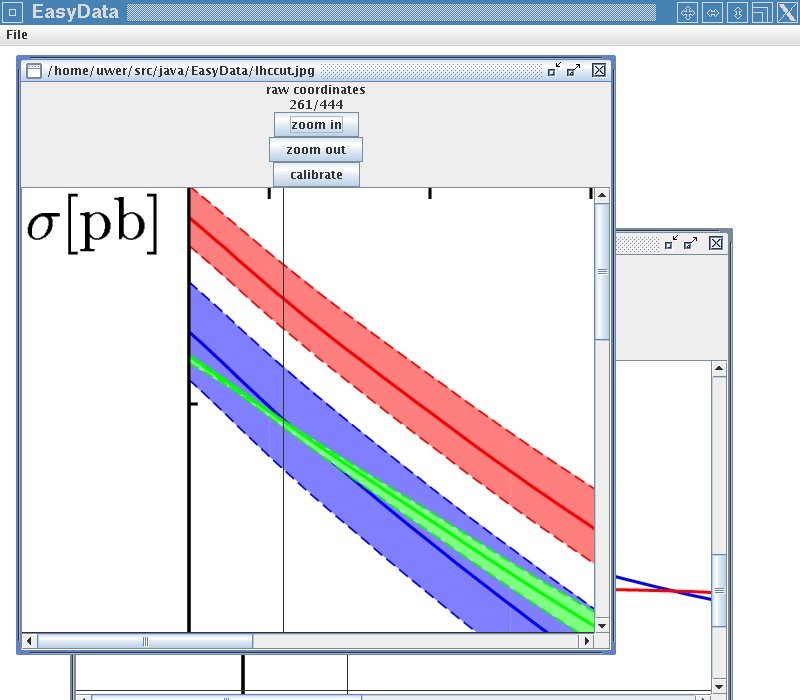}    
    \caption{Loaded plot is displayed in \EasyNData.}
    \label{fig:1}
  \end{center}
\end{figure}
\begin{figure}[htbp]
  \begin{center}
    \leavevmode
\includegraphics[width=10cm]{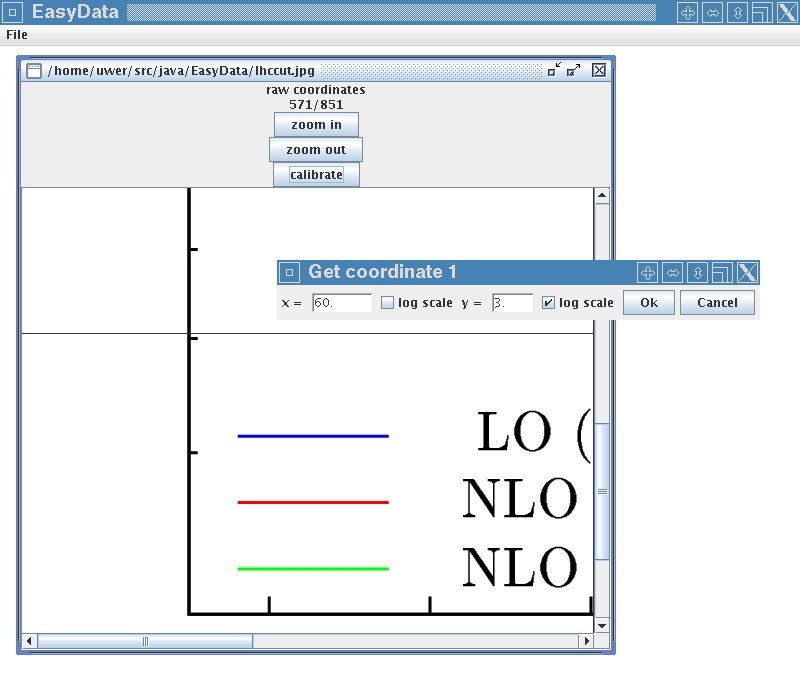}    
    \caption{Calibration procedure.}
    \label{fig:2}
  \end{center}
\end{figure}
Using \EasyNData\ the procedure is as follows:
\begin{enumerate}
\item Convert the plot to JPEG format. This can be done for example
  by a screen shot or by retrieving the Postscript file of the plot from
  {\tt http://arxiv.org/}  and converting it to JPEG format 
  using publicly available
  tools. In principle a scanner
  could also do the job. (This might be useful in cases where the plot
  is only available from a journal.)
  It pays off to create a large format to reach good accuracy. 
\item Start \EasyNData\  by calling {\tt java -jar EasyNData.jar} 
  Open the JPEG file using the file menu (Fig.1). Now one should see the figure
  inside the window (see Fig. 2). 
  Note that you can load several plots at the same time.
\item Press the button "calibrate" to start the calibration procedure.
  One should chose a point in the plot where it is easy to read off the
  coordinates. After clicking on the point a new window opens where
  the coordinates in units used in the plot have to be entered (Fig.3).
  If the corresponding axis has a logarithmic  scale the check box log scale 
  must be selected.
\item This step has to be repeated with a second point. Having done
  that the calibration is finished.
  In the first line "raw coordinates" of the plot window should now appear 
  replaced by "calibrated
  coordinates". To check that nothing went wrong one may try a few points
  where an easy extraction is possible. Note that the calibration
  procedure can be redone at any time.
\item Place the crosshair cursor where ever you like to get the
  data. The corresponding values are displayed on the fly.
\end{enumerate}
{\bf Requirements:}
The program should run on every platform where a recent runtime version of the
Java\texttrademark\ virtual machine is available. 
Note however that due to major incompatibilities between java versions
$> 1.5$ and earlier ones the version has to be at least 1.5.
In the example described below the
version used was (output from {\tt java -version}):
\begin{verbatim}
java version "1.5.0_08"
Java(TM) 2 Runtime Environment, Standard Edition (build 1.5.0_08-b03)
Java HotSpot(TM) 64-Bit Server VM (build 1.5.0_08-b03, mixed mode)
\end{verbatim}
The program was also sucessfully tested under 32bit linux
and Microsoft\textregistered\  Windows Vista\texttrademark. 
Note that the program is delivered as jar-file. No compilation is necessary.

\section{Examples}
Using \EasyNData\ I find for the upper blue line in Fig. 4 of 
ref. \cite{Dittmaier:2007th} the 
  results shown in table \ref{tab:tev}. (The upper line was chosen
  because of its steepness---any uncertainty in the $x$ determination
  adds as an important uncertainty to the $y$ determination. So the
  steeper the more difficult is the extraction.) 
  \begin{table}[htbp]
    \begin{center}
      \leavevmode
  \begin{tabular}[h]{l|c|c}
    scale $\mu/m_w$ & value obtained using \EasyNData & original data
    \cite{Dittmaier:2007th}\\ \hline
    0.1 & 5.480 & 5.50573...\\
    1.0 & 2.293 & 2.30282...\\
    3.1622 & 1.650 & 1.65366... \\
    10.0 & 1.239 & 1.24081...\\
  \end{tabular}
  \caption{Extraction of data from Fig. 4 of 
    ref. \cite{Dittmaier:2007th}}
  \label{tab:tev}
\end{center}
\end{table}
Two remarks are in order.
\begin{enumerate}
\item Although the numbers in the last
  column are given to 6 digits it is not claimed that the cross
  section is known with that precision. I just quoted the numbers as they
  were obtained from a numerical phase space integration and 
  used in the preparation of the plots.
\item In fact the extraction of the plot data consists of two
  informations. The $x$- as well as the $y$-axis have to be
  extracted. In the example the $x$ values were chosen such to match 
the sampling points
  used in the numerics. In case of the value 3.1622 it was not
  possible to exactly hit this point with the mouse due to the discret 
  sampling of the mouse position. 
  This could be solved by further increasing the bitmap or
  switching to a screen with a higher resolution. (The example was run
  on a notebook with XGA resolution that is 1024x758.)
\end{enumerate}
One might argue that this was an easy example because the $y$-axis is
linear. So let's do a second example and take the middle line of the 
red band  shown in the lower plot of Fig.6 of ref. \cite{Dittmaier:2007th}. 
The results are shown in
table \ref{tab:2}.
\begin{table}[htbp]
  \begin{center}
    \leavevmode
    \begin{tabular}[h]{l|c|c}
    {\rm cut} & value obtained using \EasyNData & original data
    \cite{Dittmaier:2007th}\\ \hline
      50 & 31.78 & 31.9544...\\ \hline
      100 & 14.65 & 14.7153...\\ \hline
      200 & 4.490 & 4.48797...\\ \hline
    \end{tabular}
    \caption{Extraction of data from Fig.6 of ref. \cite{Dittmaier:2007th}}
    \label{tab:2}
  \end{center}
\end{table}
Although the accuracy is not as good as in the first example it is
still better than a 1\%. Given that the uncertainty of the original
data points due to the numerical integration is probably of the same
size the extracted data would still allow what I would call 
a detailed comparison.

The JPEG files used in the example were created from the plots
available on {\tt http://arxiv.org/}. They are bundled together with
the program so that everybody can repeat the example.

\section{Known limitations and possible extensions}
It seems to be that zooming into  an already large image creates a
runtime exception caused by insufficient memory of the java virtual
machine. This is not a severe problem because in any case one will not
gain anything from further zooming. The main purpose of the zoom-in
button is to undo a previous zoom-out action. It might be as well that
this problem could be avoid by attributing more memory to the virtual
machine when the program is launched. For heavy use it might be
useful to dump the extracted points directly to a file. This might be 
considered in a future version. Using more than 2 points to calibrate 
the plot could further improve the quality. 
Please note that no warranty for the 
correctness of the program is given and that every user is using the
program on its own risk. 
%% Furthermore this software is not designed or 
%% intended for use in on-line control of aircraft, air traffic, 
%% aircraft navigation or aircraft communications; or in
%% the design, construction, operation or maintenance of any nuclear
%% facility.

\section{Conclusions}
In this article a short Java\texttrademark\
tool is described to extract numerical data from
published plots. Depending on the quality of the plots an accuracy at
a few per mille level can be reached. Given that in many cases the
numerical data is not more precise this is
sufficient for many detailed comparisons. I believe that statements as
made for example in ref. \cite{ellisetal} 
footnote 1 on page 2 are not state of the art, 
given that the main question is whether one agrees or not on the
prediction for measurable 
quantities. This is in particular true taken into account
that we all know that the
tricky part of one-loop corrections for the LHC is to get it
right in the entire relevant phase space (in a finite time!) and
not just in one point.

Acknowledgment: \\
I would like to thank Nigel Glover and Giampierro Passarino for a
careful reading of the manuscript and interesting discussion.

\end{document}